\documentclass[onecolumn]{sdl}

\def\0{\mbox{\tiny $0$}}
\def\1{\mbox{\tiny $1$}}
\def\2{\mbox{\tiny $2$}}
\def\3{\mbox{\tiny $3$}}
\def\4{\mbox{\tiny $4$}}
\def\5{\mbox{\tiny $5$}}
\def\6{\mbox{\tiny $6$}}
\def\7{\mbox{\tiny $7$}}
\def\8{\mbox{\tiny $8$}}
\def\9{\mbox{\tiny $9$}}

\logo{
\colorbox{DarkGoldenrod}{\color{white}$\mathbf{\Sigma\hspace*{0.06cm} \delta \hspace*{0.04cm} \Lambda}$}
}

\journal{\shadowtext{\textbf{\color{DarkRed} The European Physics Journal Plus}}\,\,\, \textbf{135}, 733-12 (2020).}

\titlelines{1}
\title{The quaternionic Goos-H\"anchen shift}

\imgbgabstract{We investigate the lateral displacement of electronic waves when partially and totally reflected by a quaternionic potential. Following the analogy between Quantum Mechanics and Optics we introduce a refractive index  for the complex and the pure quaternionic case. For incidence greater than the critical one (total reflection)  the quaternionic potentials amplify the lateral displacement found in the complex case. For incidence below the critical angle, we find, in the quaternionic case, an additional shift. The analytical formula found for the quaternionic lateral displacement gives the possibility to observe  quantitative and qualitative differences  between the complex and quaternionic Goos-H\"anchen shift.}

\author{
\names{Stefano De Leo$^{^{(1,a)}}$ and\,\, Gisele Ducati$^{^{\,(2,b)}}$}
\affiliation{$^{(1)}$ Department of Applied Mathematics, State of University of Campinas, Brazil.}
\affiliation{$^{(2)}$Center of Mathematics, Computation and Cognition, Federal University of ABC, Brazil.}
\email{$^{(a)}$ deleo@ime.unicamp.br,\,\,\,\,\,\,\,\,   $^{(b)}$ ducati@ufabc.edu.br.}
}

\begin{document}

\sdlmaketitle

\section{Introduction}

The Goos-H\"anchen effect\cite{GH47,Art48,GH49} is a well-known optical phenomenon  
regarding the lateral displacement that an optical beam of a finite cross section undergoes when it is totally reflected. In recent years,  interesting applications  are found  in different areas like slow light processes \cite{SL1,SL2}, nanophotonics \cite{NP1,NP2,NP3}, photonic crystals \cite{PC1,PC2,PC3}, and seisimc waves \cite{DK18}. Experimental evidences  of lateral displacement, angular deviations, and oscillatory phenomena due to the Goos-H\"anchen shift are  reported in \cite{Exp1,Exp2,Exp3}. For theoretical detailed reviews, we refer the reader to the tutorials given in refs. \cite{A13,M19}.

Quantitative and qualitative differences between complex and quaternionic Quantum Mechanics \cite{ADL}
had always been the main subject matter of many papers \cite{del0,del1,del2,del3,del4,del5,a08,a09,del6,del7,a14,del8,a16,a17,a18,del9}. Such investigations  surely  contribute to motivate   the worldwide community interested in the study of quaternionic formulation of Classical and Quantum Theories. In particular, a recent study on the analogy between quaternionic Quantum Mechanics and Optics \cite{SQP} led to a quaternionic formulation of the Snell law. As it  happens in the complex case, where, for total reflection, an additional phase appears in the reflection coefficient and generates the 
Goos-H\"anchen shift, also for the quaternionic case additional phases appear in the reflection coefficient and create lateral displacements.   

The paper is structured as follows. Section II presents the analogy between Quantum Mechanics and Optics and it is important to understand  the complex Goos-H\"anchen shift. 
Since  quaternionic Optics is not a known topic and the introduction of the quaternionic Snell law  is short and pretty simple, for the convenience of the reader, in Section III, we  repeat the derivation of the quaternionic Snell law, thus making our exposition self-contained.   In Section IV, we obtain the quaternionic reflection coefficient. The study of the additional phases  for the complex and the pure quaternionic case is done in Section V.  The last Section contains  our final considerations and suggestions of possible future investigations.

\section{Analogy between Optics and Quantum Mechanics}
\label{secII}

 The analogy between Optics and Quantum Mechanics \cite{cohen} allows to obtain the Snell law by analyzing the reflected and refracted waves for the  planar  $(y,z)$ motion  of a quantum mechanical particle in the presence of a stratified potential,
\begin{equation}
V(z) = \left\{\,0\,\,\,\,\,\mbox{for}\,\,\,z<0\,\,\,\,\,\,\,\mbox{and}\,\,\,\,\,\,\,
V_{\1}\,\,\,\,\,\mbox{for}\,\,\,z>0\,\right\}\,\,,
\end{equation}
whose discontinuity is for convenience of calculation fixed at $z=0$. The solution of the complex Schr\"odinger equation for $z<0$ (region I)  contains two waves, the incoming and the reflected ones, see Fig.\,1,
\begin{equation}
\psi_{_{\mathrm{I}}}(y,z) = \exp[\,i\,(p_y\,y+ p_z\, z)/\hbar\,] + r\,\exp[\,i\,(p_y\,y- p_z\,z)/\hbar\,]\,\,,
\end{equation}
where $r$ is the  reflection coefficient.  For $z>0$ (region II) the solution contains the transmitted wave, see Fig.\,1,
\begin{equation}
\psi_{_{\mathrm{II}}}(y,z) = t\,\exp[\,i\,(p_y\,y+ q_z\,z)/\hbar\,]\,\,,
\end{equation}
where $t$ is  the transmission coefficient. Note that due to the fact that the discontinuity is along the $z$-axis, the  momentum component perpendicular to this axis  is not changed when crossing the the interface.

The reflection and transmission coefficients can then be obtained by imposing the continuity constraints to the wave function and its derivative. This yields to \cite{cohen}  
\begin{equation}
\label{rc}
r = \displaystyle \frac{p_{z} - q_{z}}{p_{z} + q_{z}}\,\,\,\,\,\,\,\mbox{and}\,\,\,\,\,\,\,
t= \frac{2\,p_z}{p_{z} + q_{z}}\,\,.
\end{equation}
Region I  represents the plane zone which is potential free. Consequently, from the Schr\"odinger equation,
\begin{equation}
\label{csef}
 E\,\psi_{_{\mathit I}}(y,z) = -\,\left[\,\frac{\,\,\,\hbar^{^{2}}}{2\,m}\,\left(\,\partial_{yy}+\partial_{zz}\, \right)\,\right]\,\psi_{_{\mathit I}}(y,z)\,\,,
\end{equation}
we then obtain the well known momentum/energy relation,
\begin{equation*}
p_{y}^{^{2}}+\,p_z^{^{2}}=p^{^{2}}=2\,m\,E\,\,.
\end{equation*}
In region II, the presence of a  potential, $V_{\1}$, implies
\begin{equation}
\label{cse}
 E\,\psi_{_{\mathrm{II}}}(y,z) = -\,\left[\,\frac{\,\,\,\hbar^{^{2}}}{2\,m}\,\left(\,\partial_{yy}+\partial_{zz}\, \right)- V_{\1}\,\right]\,\psi_{_{\mathrm{II}}}(y,z)\,\,,
\end{equation}
from which we obtain the following momentum/energy relation
\begin{equation}
\label{pev1}
  p_{y}^{^{2}} + \,q_{z}^{^{2}} = 2\,m\,( E - V_{\1} ) = \left(1-\frac{V_{\1}}{E}\right)\,p^{\2}= n^{\2}p^{\2}\,\,,
\end{equation}
where we have introduced the dimensionless quantity
\begin{equation} \label{ncomp} n =  \sqrt{1-\frac{V_{\1}}{E}}\,\,,
\end{equation}
playing the role of the refractive index of Optics \cite{saleh,born}. Observing that $p_{z}=p\,\cos\theta$ and $q_{z}=n\,p\,\cos\varphi$, see Fig.\,1, from Eq.(\ref{pev1}) we find
\begin{equation*}
p^{\2}\,\sin^{\2}\theta + n^{\2}p^{\2}\,\cos \varphi^{\2} = n^{\2} p^{\2}
\end{equation*}
and consequently, we recover, for a quantum non relativistic particle crossing the interface between the free and the potential region, the well-known Snell law,  
\begin{equation}
\label{snell-c}
\sin \theta =n \,\sin \varphi\,\,.
\end{equation}

\subsection{The Goos-H\"anchen shift}

The Goos-H\"anchen effect occurs when incident angles are bigger than the critical angle,
\begin{equation}
\theta_{\mathrm{cri}}\,=\,\arcsin[\,n\,]\,=\,\arcsin\left[\,\sqrt{\frac{E\,-\,V_{\1}}{E}}\,\,\right]\,\,.
\end{equation}
 In order to  calculate the lateral displacement it is useful to rewrite the reflection coefficient in terms of the incident angle $\theta$,
\begin{equation}
\label{rcom}
r = \frac{\cos\theta-n\,\cos\varphi}{\cos\theta+n\,\cos\varphi} =
\frac{\cos\theta - \sqrt{n^{\2}-\sin^{\2}\theta}}{\cos\theta + \sqrt{n^{\2}-\sin^{\2}\theta}}\,\,.
\end{equation}
For incidence angle satisfying  the constraint $\sin\theta>n$, the reflection coefficient becomes
\begin{equation}
r = \exp\left[\,-\,2\,i\,\psi_{_{\mathrm{GH}}}\,\right] \,\,,
\end{equation}
where
\begin{equation}
\psi_{_{\mathrm{GH}}} = \arctan\left[\,\frac{\sqrt{\sin^{\2}\theta -n^{\2}}}{\cos\theta}\,\right]
\end{equation}
is the phase, utilized by Artmann \cite{Art48} to  explain the  Goos-H\"anchen shift obtained for transverse electric waves \cite{GH47}. Let us briefly recall why the presence of such a phase in the reflection coefficient implies a lateral displacement of the reflected particle. Let us first consider the incoming particle 
whose phase is given by 
\begin{equation}
\phi_{_{\mathrm{INC}}}\,=\,(\,p_y\,y_{_{\mathrm{INC}}}\,+\,p_z\,z\,)\,/\,\hbar=\,p\,\left(\,\sin \theta\, y_{_{\mathrm{INC}}} \, + \, \cos \theta\, z\,\right)\,/\,\hbar\,\,.
\end{equation}
For wave packet, the main contribution to the motion of the particle comes from the minimum of such a phase \cite{spm}. By  imposing a null derivative, we find 
\begin{equation}
\frac{\partial \phi_{_{\mathrm{INC}}}}{\partial\,\theta}\,=\,0\,\,\,\,\,\,\,\Rightarrow\,\,\,\,\,\,\, y_{_{\mathrm{INC}}}\,=\,\tan\theta\,z\,\,.
\end{equation}
For the reflected wave, when we are in the case of total reflection, the phase is given by 
\begin{equation}
\phi_{_{\mathrm{REF}}}\,=\,(\,p_y\,y_{_{\mathrm{INC}}}\,-\,p_z\,z\,)\,/\,\hbar\,-\,2\, \psi_{_{\mathrm{GH}}}  \,=\,p\,\left(\,\sin \theta\, y_{_{\mathrm{REF}}} \, - \, \cos \theta\, z\,\right)\,/\,\hbar\,-\, 2\,\psi_{_{\mathrm{GH}}}\,\,.
\end{equation}
Consequently, we find
\begin{equation}
\frac{\partial \phi_{_{\mathrm{REF}}}}{\partial\,\theta}\,=\,0\,\,\,\,\,\,\,\Rightarrow\,\,\,\,\,\,\, y_{_{\mathrm{INC}}}\,=\,-\,\tan\theta\,z\,+\,\frac{2\,\hbar}{p\,\cos\theta}\,\frac{\partial \psi_{_{\mathrm{GH}}}}{\partial\,\theta}   \,\,.
\end{equation}
Thus, the beam is reflected with an angle equal to the incident one (reflection law) but is laterally displaced. This lateral displacement, known as Goos-H\"anchen shift,  is given by
\begin{equation}
y_{_{\mathrm{GH}}}\,=\,\frac{2\,\hbar}{p\,\cos\theta}\,\frac{\partial \psi_{_{\mathrm{GH}}}}{\partial\,\theta}\,=\,
\,\frac{2\,\hbar \, \tan\theta}{p\,\sqrt{\sin^{\2}\theta -n^{^2}}}\,\,.
\end{equation}
In Fig.\,1, we illustrate this phenomenon for a potential $V_{\1}=E/2$. In this case, the critical angle is $\pi/4$. We conclude this Section, by oberving that the infinity for critical incidence is removed by using integral Fourier transforms \cite{inf1,inf2,inf3}.

\section{The quaternionic Snell law}

In the previous Section, we have obtained the Snell law and the reflection coefficient for planar motion in the presence of complex potentials. In this Section, we show how  the Snell law is modified when 
we use a quaternionic potential in the Schr\"odinger equation.  The time-independent Schr\"odinger equation in the presence of a quaternionic potential,
\[ \boldsymbol{h}\cdot \boldsymbol{V}(z) =\left\{\,0\,\,\,\,\,\mbox{for}\,\,\,z<0\,\,\,\,\,\,\,\mbox{and}\,\,\,\,\,\,\,
i\,V_{\1}+j\,V_{\2}+k\,V_{\3}\,\,\,\,\,\mbox{for}\,\,\,z>0\,\right\}\,\,,\]
is given by \cite{ADL}
\begin{equation}
\label{qse}
E\,\Psi_{_{II}}(y,z)\,i = \left[\,-\,i\,\frac{\,\,\,\hbar^{^{2}}}{2\,m}\,\left(\,\partial_{yy}+\partial_{zz}\, \right)\,+\,  \boldsymbol{h}\cdot \boldsymbol{V}\,\right]\,\Psi_{_{II}}(y,z)
 = A_{_{\mathrm{H}}}\,\,\Psi_{_{II}}(y,z)\,\,.
\end{equation}
Multiplying the previous equation from the left by the quaternionic anti-hermitian operator 
$A_{_{\mathrm{H}}}$, we get
\begin{eqnarray*}
E^{^{2}}\,\Psi_{_{II}}(y,z)  & = &  -\,A_{_{\mathrm{H}}}^{^{2}}\,\,\Psi_{_{II}}(y,z)\\
 & = &\left[\,\frac{\,\,\,\hbar^{^{4}}}{4\,m^{\2}}\,\left(\,\partial_{yy}+\partial_{zz}\, \right)^{^{2}} \,-\, \frac{\,\,\hbar^{^{2}}}{m}\,\,V_{\1}\,\, \left(\,\partial_{yy}+\partial_{zz}\, \right)   \,+\, |\boldsymbol{V}|^{^{2}}\,\right]\,\Psi_{_{II}}(y,z)\,\,.
\end{eqnarray*}
This equation allows to  obtain the momentum solution which represents the quaternionic counterpart of  $q_{z}$, i.e.
\begin{equation}
 \label{quat1}
 p_{y}^{^{2}} + Q_{z}^{^{2}} = 2\,m\,\left(\,\sqrt{E^{^{2}}-V_{\2}^{^{2}}-V_{\3}^{^{2}}} - V_{\1}\,\right)\,\,,
\end{equation}
which characterizes  the plane wave in the potential region,
\[  \exp[\,i\,(\,p_{y}\,y+ Q_{z}\,z)\,/\hbar\,]\,\,, \]
and the additional solution
\begin{equation}
 \label{quat2}
 p_{y}^{^{2}} + \widetilde{Q}_{z}^{^{\,2}} = -\,
 2\,m\,\left(\,\sqrt{E^{^{2}}-V_{\2}^{^{2}}-V_{\3}^{^{2}}} + V_{\1}\,\right)
\end{equation}
which generates {\em evanescent} wave solutions,
\[  \exp[\,(\,i\,p_{y}\,y - |\widetilde{Q}_{z}|\,z)\,/\hbar\,]\,\,. \]
Before giving the explicit quaternionic solutions  for the reflection coefficient and compare them with the complex case, let us now examine how the Snell law is modified by the {\em new} momentum $Q_{z}$. From Eq.(\ref{quat1}), we find
\begin{equation}
\label{pev2}
  p_{y}^{^{2}} + \,Q_{z}^{^{2}} = \left(\,\sqrt{1-\frac{V_{\2}^{^{2}}+V_{\3}^{^{2}}}{E^{^{2}}}} - \frac{V_{\1}}{E}\,\right)\,2\,m\,E =
N^{^2}p^{\2}\,\,.
\end{equation}
The presence of a quaternionic part in our potential generates the refractive index
\begin{equation}  \label{Nquat}
N = \sqrt{\sqrt{1-\frac{V_{\2}^{^{2}}+V_{\3}^{^{2}}}{E^{^{2}}}} - \frac{V_{\1}}{E}}
\end{equation}
and consequently the {\em new} Snell law
\begin{equation}
\sin \theta = N\, \sin \phi\,\,.
\end{equation}

\section{The reflection amplitudes for quaternionic potentials}

To obtain the reflection amplitude, we have to impose the continuity
of the wave function and its derivative at the potential discontinuity $z=0$. The quaternionic plane wave solutions are given  by \cite{del0,del1,del2,del3,del4,del5}
\begin{equation}
\Psi_{_{\mathit I}}(y,z) =  \left\{  \exp[i\,p_zz]  + R\, \exp[-\,i\,p_{z}z] + j\, \widetilde{R}\,\exp[p_{z}z]   \right\}\,  \exp[i\,p_{y}y]
\end{equation}
in the free potential region ($z<0$) and 
\begin{equation}
\Psi_{_{\mathit II}}(y,z) =  \left\{ (1+j\,\beta)\, T\, \exp[\,i\,Q_{z}z] + (\alpha +j\,)\,\widetilde{T}\, \exp[\,-\,|\widetilde{Q}_{z}|\,z]   \right\}\,  \exp[i\,p_{y}y]\,\,,
\end{equation}
where
\[ \alpha= i\,\frac{(V_{\2}+i\,V_{\3})/E}{1+N^{^{2}}}\,\,,\,\,\,\,\,\,\,\,\,\,
\beta= -\,i\,\frac{(V_{\2}-i\,V_{\3})/E}{1+N^{^{2}}}\,\,,
\]
in the potential region ($z>0$). From the continuity equations of the pure quaternionic wave functions and their derivative, we get
\begin{equation}
\widetilde{R}  =    \beta\, T + \widetilde{T}\,\,\,\,\,\,\,\,\,\,\mathrm{and}\,\,\,\,\,\,\,\,\,\,
p_{z}\,\widetilde{R} =   i\, \beta\, T\,Q_{z} - \widetilde{T}\,|\widetilde{Q}_{z}|\,\,,
\end{equation}
from which we obtain
\[ \widetilde{T} =\frac{i\,Q_{z} - p_z}{|\widetilde{Q}_{z}| + p_{z}}\,\,\beta\,  T\,\,.\]
From the continuity equations of the complex wave functions and their derivative, we get
\begin{equation}
   1+ R =    T+\alpha\,  \widetilde{T}\,\,\,\,\,\,\,\,\,\,\mbox{and}\,\,\,\,\,\,\,\,\,\,
   1 -  R  =  \displaystyle{\frac{Q_{z}}{p_{z}}}\,T + i\,\displaystyle{\frac{|\widetilde{Q}_{z}|}{p_{z}}}\,\,\alpha\,  \widetilde{T}\,\,,
\end{equation}
from which we obtain
\[
\frac{1+R}{1-R}=\frac{p_z+\alpha\,p_z \widetilde{T}/T}{Q_z+i\,\alpha\,|\widetilde{Q}_z | \,\widetilde{T}/T  }\,\,.
\]
Finally,
\begin{eqnarray}
\label{rquat}
R&=&\frac{(\,p_{z}\,-\, Q_{z}\,)(\,|\widetilde{Q}_{z}|\, +\, p_{z}\,)\,+\,\alpha\,\beta\,(\,i\,Q_{z}\,-\,p_z)(\,p_{z} \,-\,i\,| \widetilde{Q}_{z}|\,)}{(\,p_{z}\,+\, Q_{z})(\, |\widetilde{Q}_{z}| \,+\, p_{z}\,)\,+\,\alpha\,\beta\,(\,i\,Q_{z}\,-\,p_z\,)(\,p_{z}\, +\,i\,|\widetilde{Q}_{z}|\,)}~.
\end{eqnarray}

\section{The quaternionic Goos-H\"anchen shift}

In this Section, we study the Goos-H\"anchen shift in the presence of a pure quaternionic potential.
To compare the quaternionic shift with the standard complex one we shall use the same refractive index for the pure quaternionic and complex potentials, i.e.
\begin{equation}
n^{\2} \,=\, 1\,-\,\frac{V_{\1}}{E}\,=\, \sqrt{ 1 - \frac{V_{\2}^{^2}+\,V_{^3}^{^2}}{E^{^2}}}~.
\end{equation}
This means that, for example, a complex potential with $V_{\1}=E/2$ and a pure quaternionic potential with 
$\sqrt{V^{^2}+\,V_{^3}^{^2}}=\sqrt{3}\,E/2$ generates the same critical angle, $\theta_{\mathrm{cri}}=\pi/4$, see Fig.\,2.

Observing that, for a pure quaternionic potential, 
\[\alpha\,\beta\,=\, \frac{1\,-\,n^{{\2}}}{1\,+\,n^{{\2}}}\,\,,\]
the reflection coefficient of Eq.\,(\ref{rquat}) can be rewritten as follows
\begin{equation}
\label{rquat2}
R\,=\,\frac{(\,1\,+\,n^{{\2}}\,)\,(\,p_{z}\,- \,Q_{z}\,)\,(\, |\widetilde{Q}_{z}| \,+\, p_{z}\,)\,+\,
(\,1\,-\,n^{{\2}}\,)\,(\,i\,Q_{z}\,-\,p_z\,)\,(\,p_{z} \,-\,i\, |\widetilde{Q}_{z}|\,)}{
(\,1\,+\,n^{{\2}}\,)\,
(\,p_{z}\,+\, Q_{z}\,)\,(\, |\widetilde{Q}_{z}| \,+\, p_{z}\,)\,+\,
(\,1\,-\,n^{{\2}}\,)
\,(\,\,i\,Q_{z}\,-\,p_z\,)\,(\,p_{z}\, +\, i\, |\widetilde{Q}_{z}|\,)}\,\,.
\end{equation}
where  
\[
\left\{\,p_z\,,\,Q_z\,,\, |\widetilde{Q}_{z}|\,\right\}\,=\,
p\,\left\{\,\cos\theta\,,\,\sqrt{n^{^{2}}\,-\,\sin^{\2}\theta}\,,\,
\sqrt{n^{^{2}}\,+\,\sin^{\2}\theta}\,\right\}\,\,.
\]
The reflection coefficients for complex and pure quaternionic potentials generating the same refractive index are plotted in Fig.\,3, for refractive indexes  corresponding to critical angles of $\pi/6$ (a), $\pi/4$ (b), and $\pi/3$ (c).  or $\sin\theta>n$, we have 
\[Q_{z}=i\,|Q_z |= i\,p\,\sqrt{\sin^{\2}\theta-n^{^{2}}}\]
and Eq.\,(\ref{rquat2}) becomes
\begin{equation}
\label{rquat3}
R_{_{>}}\,=\,\frac{(\,1\,+\,n^{^{2}}\,)\,(\,p_{z}\,- \,i\,|Q_{z}|\,)\,(\, |\widetilde{Q}_{z}| \,+\, p_{z}\,)\,-\,
(\,1\,-\,n^{^{2}}\,)\,(\,p_{z}\,+\,|Q_{z}|\,)\,(\,p_{z} \,-\,i\, |\widetilde{Q}_{z}|\,)}{
(\,1\,+\,n^{^{2}}\,)\,
(\,p_{z}\,+\,i\, |Q_{z}|\,)\,(\, |\widetilde{Q}_{z}| \,+\, p_{z}\,)\,-\,
(\,1\,-\,n^{^{2}}\,)
\,(\,p_{z}\,+\,|Q_{z}|\,)\,(\,p_{z}\, +\,i\,|\widetilde{Q}_{z}|\,)}\,\,.
\end{equation}
As happens for the complex case, the particle is totally reflected, see Fig.\,3, 
and the  reflection coefficient can be written in the following form  
\begin{equation}
R_{_{>}}\,=\, \exp[\,-\,2\,i\,\Psi_{_{\mathrm{GH}}}]
\end{equation}
where
\begin{equation}
\tan \Psi_{_{\mathrm{GH}}}\,=\, \frac{
(\,1\,+\,N^{^{2}}\,)\,|Q_z|\,\,(\,p_z \,+\,|\widetilde{Q}_z| \,)\,-\,
(\,1\,-\,N^{^{2}}\,)\,|\widetilde{Q}_z|\,\,(\,p_z \,+\,|Q_z| \,)}{
(\,1\,+\,N^{^{2}}\,)\,\,p_z\,(\,p_z \,+\,|\widetilde{Q}_z| \,)\,-\,
(\,1\,-\,N^{^{2}}\,)\,\,p_z\,\,(\,p_z \,+\,|Q_z| \,)
}
\end{equation}
The  stationary phase method yields
\begin{equation}
y^{^{\mathrm{[quat]}}}_{_{\mathrm{GH}}} \,=\,\frac{2\,\hbar}{p\,\cos\theta}\,\frac{\partial \Psi_{_{\mathrm{GH}}}}{\partial\,\theta}~.
\end{equation}
In Fig.\, 4, the adimensional Goos-H\"anchen shift $p\,y_{_{\mathrm{GH}}}/\hbar$ is plotted for complex and pure quaternionic potentials. The quantitative differences appear for incidence angles greater than the critical one. 

Qualitative differences appear for incidence below the critical angle, i.e. 
$\sin\theta<n$. In this case, differently of what happens for the complex case where the reflection coefficient is real, the reflection coefficient is still a complex number,
\begin{equation}
R_{_{<}}\,=\, \left|\, R_{_{<}}\,\right|\,\exp[\,i\,(\,\Phi_{_{\mathrm{GH}}}^{^{(\mathrm{num})}}\,-\,
\Phi_{_{\mathrm{GH}}}^{^{(\mathrm{den})}}\,  ]\,\,,
\end{equation}
with
\begin{equation}
\tan\left[\,\Phi_{_{\mathrm{GH}}}^{^{(\mathrm{num})}}\,\right] \, = \, \frac{(\,1\,-\,n^{^{2}}\,)\,
(\, Q_z\,+\,|\,\widetilde{Q}_z|\,)\,p_z}{(\,1\,+\,n^{^{2}}\,)\,(\,p_{z}\,- \,Q_{z}\,)\,(\, |\widetilde{Q}_{z}| \,+\, p_{z}\,)\,-\,
(\,1\,-\,n^{^{2}}\,)\,(\,p^{^{2}}_{z}\,-\,Q_{z}\, |\widetilde{Q}_{z}|\,)}
\end{equation}
and
\begin{equation}
\tan\left[\,\Phi_{_{\mathrm{GH}}}^{^{(\mathrm{den})}}\,\right] \, = \, \frac{(\,1\,-\,n^{^{2}}\,)\,
(\, Q_z\,-\,|\,\widetilde{Q}_z|\,)\,p_z}{(\,1\,+\,n^{^{2}}\,)\,(\,p_{z}\,+ \,Q_{z}\,)\,(\, |\widetilde{Q}_{z}| \,+\, p_{z}\,)\,-\,
(\,1\,-\,n^{^{2}}\,)\,(\,p^{^{2}}_{z}\,+\,Q_{z}\, |\widetilde{Q}_{z}|\,)}\,\,.
\end{equation}
In the pure quaternionic case, a lateral displacement is thus present also for incidence below the critical angle,
\begin{equation}
\widetilde{y}^{^{\,\mathrm{[quat]}}}_{_{\mathrm{GH}}} \,=\,\frac{\hbar}{p\,\cos\theta}\,\left[\,\frac{\partial \Phi_{_{\mathrm{GH}}}^{^{(\mathrm{den})}}}{\partial\,\theta}\,-\, \frac{\partial \Phi_{_{\mathrm{GH}}}^{^{(\mathrm{den})}}}{\partial\,\theta}\,\right]
~.
\end{equation}
This additional Goos-H\"anchen shift is shown in the plots of Fig.\,4.

\section{Conclusions}
Quantum Mechanics works very well and  Field Theory complements  the prediction of Quantum Mechanics in the case of creation and annihilation of particles.    However, this does not mean that it is impossible, in principle, to observe deviations from the complex Quantum Mechanics.  Adler was the first to write a book on quaternionic Quantum Mechanics \cite{ADL} and, starting from this milestone work, many papers investigated the possibility to find quantitative and qualitative differences between the complex and the quaternionic formulation of Quantum Mechanics. The analogy between Optics and Quantum Mechanics allows to propose, for electronic waves, a well-known phenomenon in Optics known as Goos-H\"anchen shift, i.e. the lateral displacement of light when it is totally reflected by the interface between two media with different refractive indexes (the first one greater than the second one). In this paper, we have compared the shift of electronic waves when reflected by complex and pure quaternionic potentials. The analysis was done for complex and quaternionic potential generating the same  critical angles. Analytical expression was obtained for the lateral displacement of electronic wave in the presence of pure quaternionic potentials. In the quaternionic case, the shift for incidence angles greater than the critical one amplifies the effect found in the complex case and presents a new phenomenon for incidence below the critical angle. The study was done by using plane waves. A future proposed work could be the analysis of the phenomena presented in this paper by using wave number distribution. This is usually done in the complex case \cite{inf1,inf2}  and it is fundamental to remove the infinity at the critical incidence and the discontinuity between the left and right regions, see Fig.\,4.  In the left region should appear the phenomenon of angular deviations and in the right region the pure Goos-H\"anchen shift \cite{M19}. The critical region should be characterized by the composite Goos-H\"anchen shift \cite{Exp1,Exp3,inf3}.

In this paper, we have studied in detail the case of a pure quaternionic potential. The most general quanternionic potential contains an additional complex part reproducing  the standard predictions of complex Quantum Mechanics. The pure complex part leading to a lateral shift for incidence greater than the critical one and the real part responsible for the absorption effect. In this case the pure quaternionic part has to be treated as a perturbation with respect to the complex one.      

The possibility to propose an experiment to confirm or rule out quaternionic formulation of Quantum Mechanics was one of the main interests in this field, see for example the first proposals of ref.\,\cite{expq1,expq2} investigating the effect of non-commutative phases and the recent one using metamaterial \cite{nature}. The main contribution of this article was the  lateral displacements caused by the pure quaternionic before the critical angle. This effect, as well as the standard Goos-H\"anchen shift, is of the order of the wavelength of the incident field. The idea for a possible proposal in testing quaternionic theories comes for the strict analogy between Quantum Mechanics and Optics. The study presented in this article has an immediate counterpart in the lateral shift of  Transverse Electric waves. Material with a pure quaternionic refractive index should reproduce the additional lateral shift before the critical angle. It is also interesting to observe that extending the study of pure quaternionic material to Transverse Magnetic waves, we could us the Weak Measurements interference technique\,\cite{wm} to amplify such lateral displacements.  I this view, the present work represents a preliminary study to stimulate future investigations in Quantum Mechanics as well in Optics.

\subsection*{Acknowledgements}

The authors thank an anonymous referee for his comments and suggestions. In particular, the authors appreciated very much the observations of the referee which have stimulated the discussion presented at the end of the concluding section. One of the authors (SdL) acknowledges the financial support from the CNPq (grant number 2018/03911) and the Fapesp (grant number 2019/06382-9).

\newpage

\vspace*{3cm}

\WideFigureSideCaption{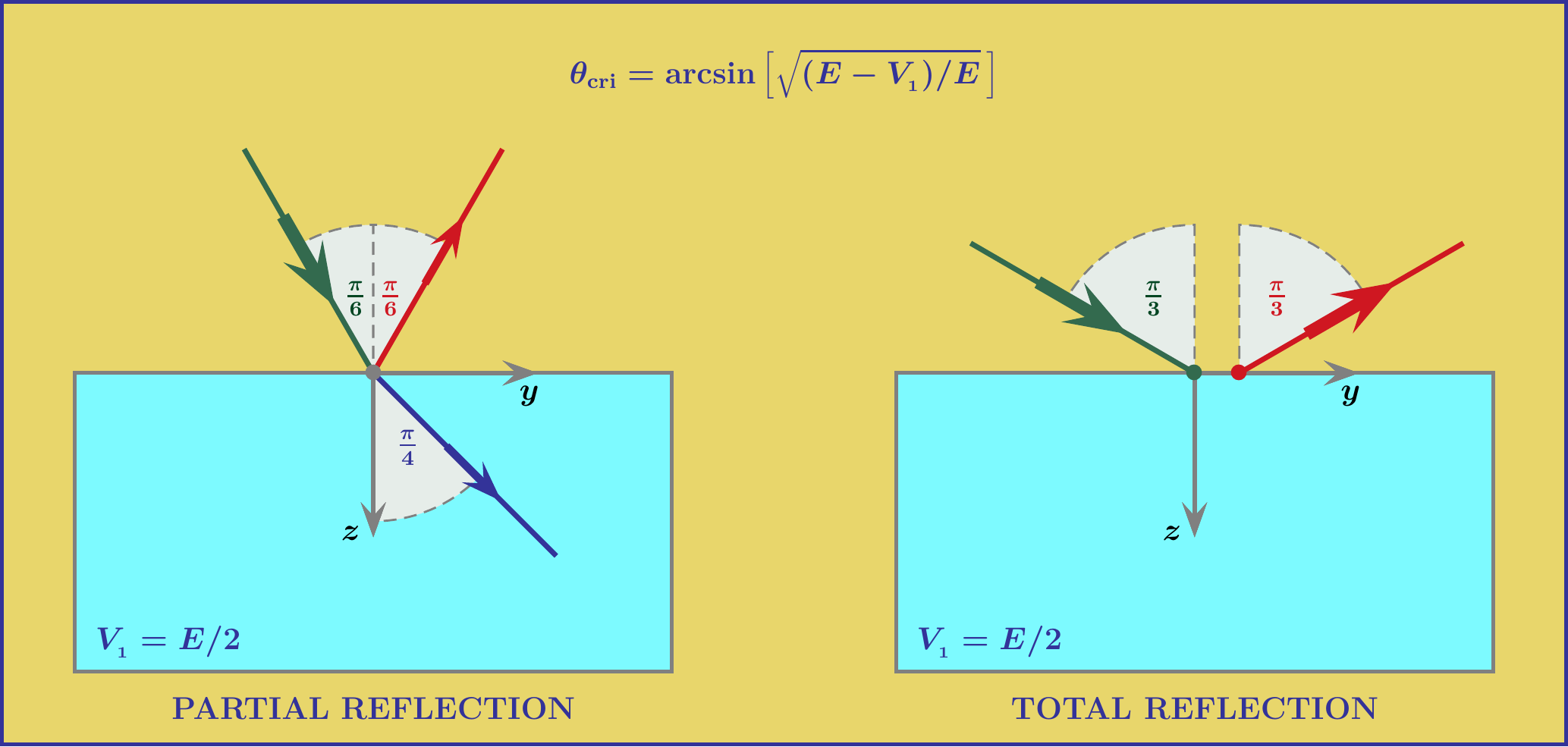}{\textbf{The complex Goos-H\"anchen shift.} For the electronic reflected wave 
a lateral displacement is found for incidence angles greater than the critical one (total reflection). For incoming particles with energy $E$ equal to $2\,V_{\1}$ the critical angle is $\pi/4$. }

\vspace*{2cm}

\WideFigureSideCaption{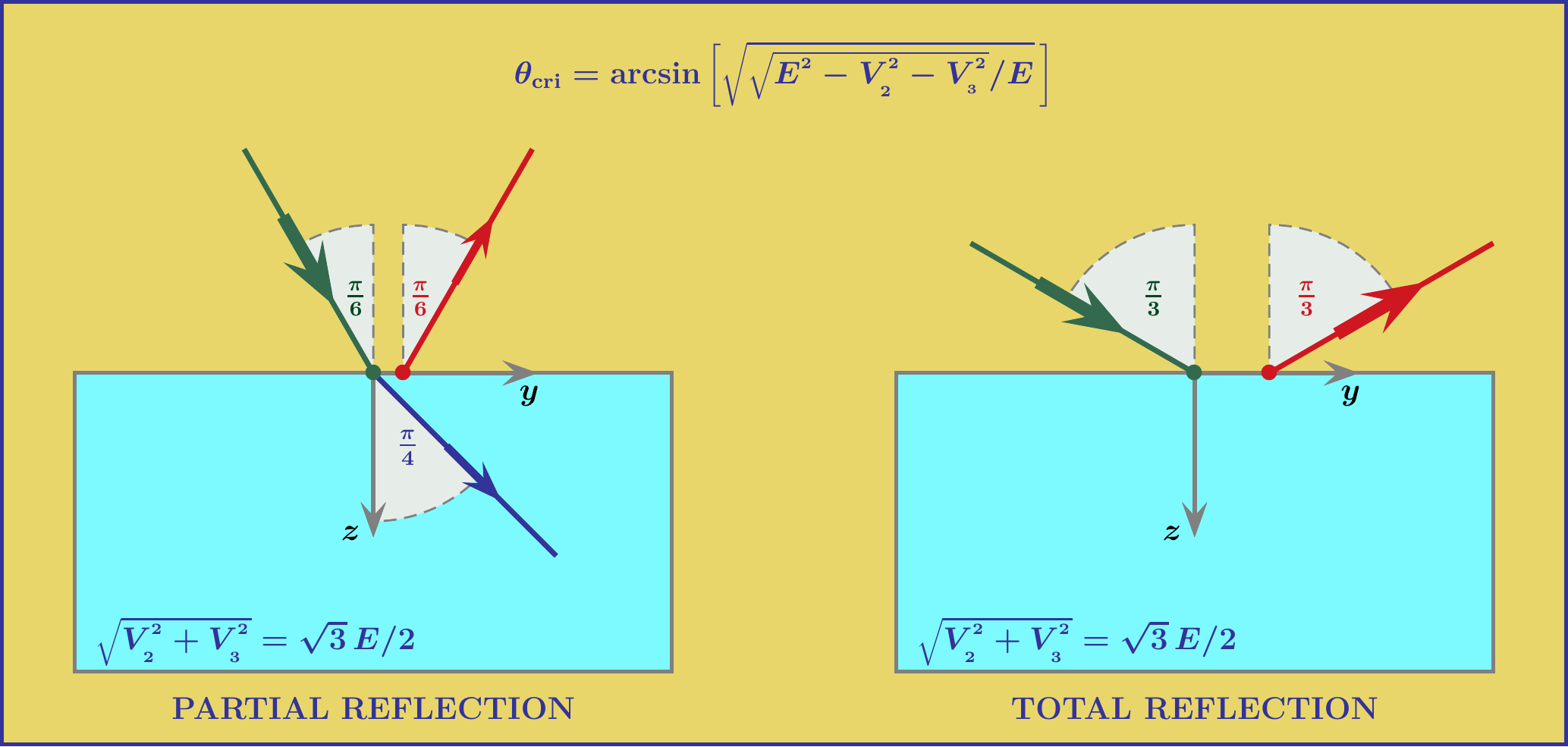}{\textbf{The quaternionic Goos-H\"anchen shift.} In the case of a  pure quaternionic potential and an incoming electronic wave with energy $E$ equal to $2\,\sqrt{\left(V_{\2}^{^{2}}+V_{\3}^{^{2}}\right)\,/\,3\,}$ the critical angle is $\pi/4$, as the complex case of Fig.\,1. For pure quaternionic potential a lateral displacement is also seen for incidence angles below the critical one.}

\newpage

\WideFigure{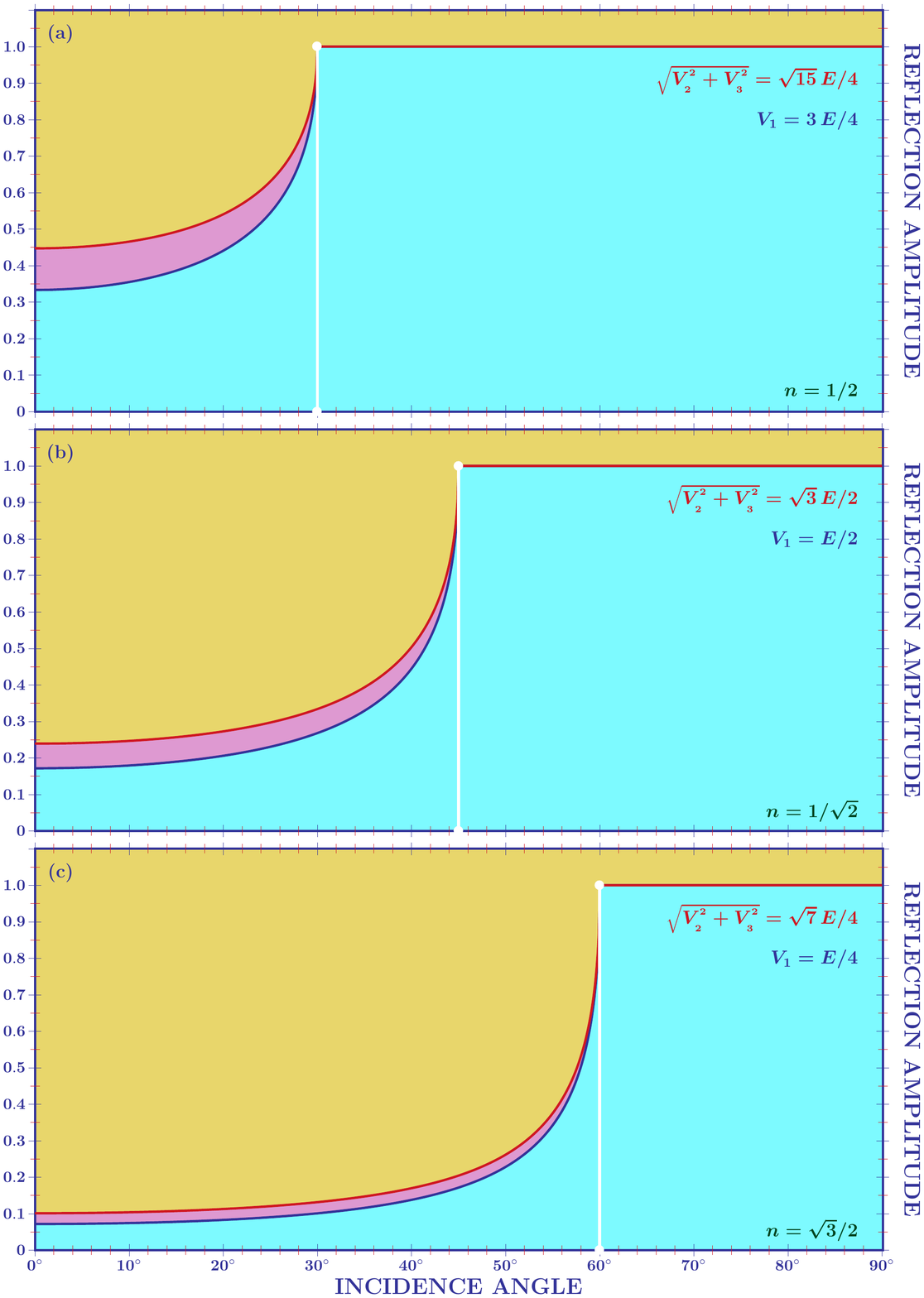}{\textbf{Reflection amplitudes.} The complex (blue line), $|r|$, and quaternionic (red line), $|R|$, reflection amplitudes  are plotted as functions of the incident angle for different potentials chosen in order to obtain the same critical angle for the complex and quaternionic case. In (a) the critical angle is $\pi/6$, in (b) $\pi/4$, and in  (c) $\pi/3$.}

\newpage

\WideFigure{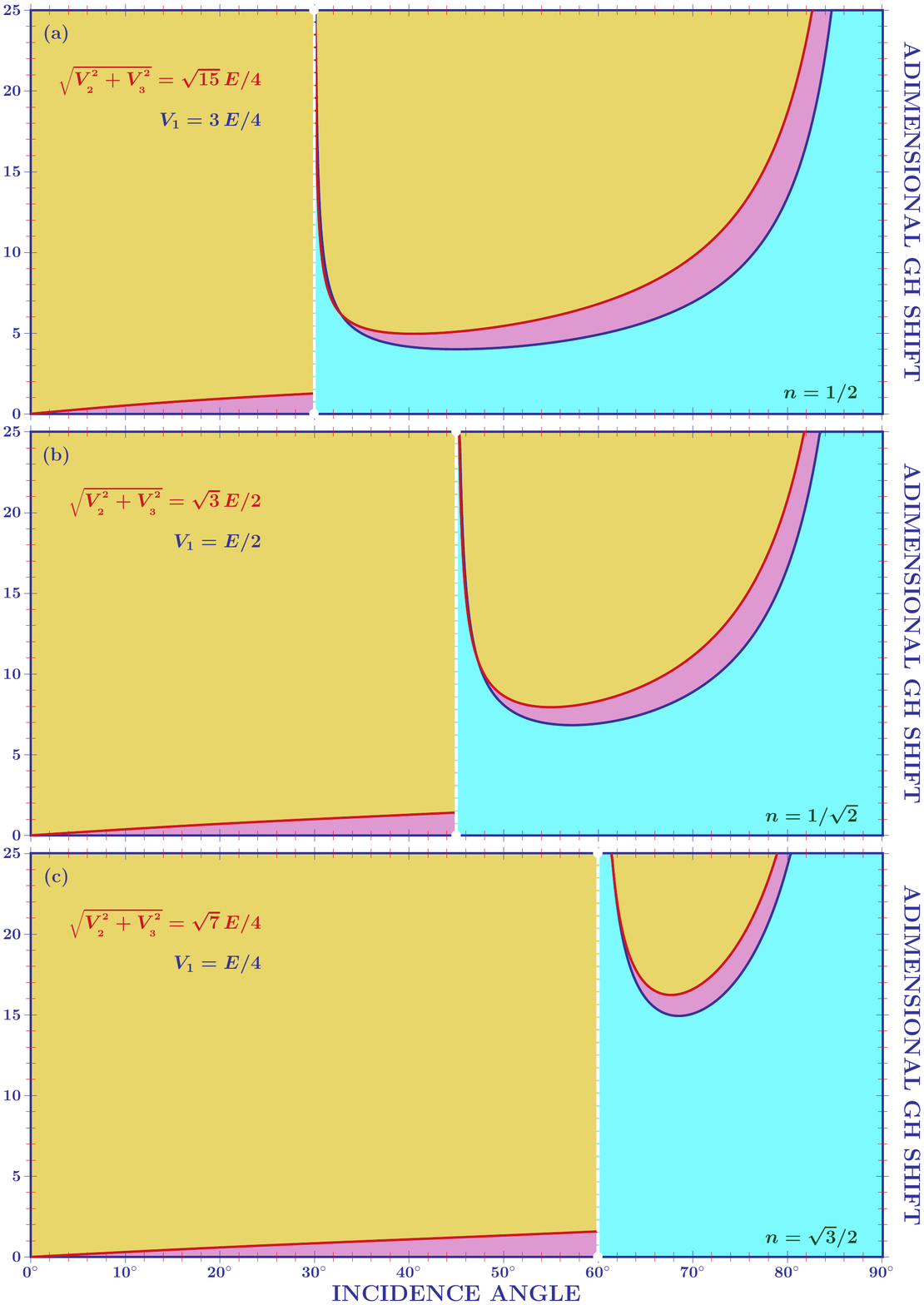}{\textbf{Goos-H\"anchen shifts.}. The adimensional lateral displacement, 
$p\,y_{_{\mathrm{GH}}}\,/\,\hbar$, is plotted for complex (blue line) and pure quaternionic (red line) potentials generating the same critical angle. Quaternionic potential amplifies the effect for incidence greater than the critical one (total reflection). A new additional shift  is seen for quaternionic potentials for incidence below the critical one.}


\begin{thebibliography}{99}

\bibitem{GH47}
F. Goos and H. H\"anchen,
{\em Ein neuer und fundamentaler Versuch zur Totalreflexion},
Ann. Phys. {\textbf{436}}, 333-346 (1947).

\bibitem{Art48}
K. Artmann,
{\em Berechnung der Seitenversetzung des totalreflektierten Strahles},
Ann. Phys. {\textbf{437}}, 87-102 (1948).

\bibitem{GH49}
F. Goos and H. H\"anchen,
{\em Neumessung des Strahlwersetzungseffektes bei Totalreflexion},
 Ann. Phys. {\textbf{440}}, 251-252 (1949).



\bibitem{SL1}
\refa{Y. Hirai, K. Matsunaga, Y. Neo and T. Matsumoto}{Observation of Goos-Hänchen shift in plasmon-induced transparency}{App. Phys. Lett.}{112-6}{051101}{2018}

\bibitem{SL2}
\refa{A. Malik, Y. Chaung, M. Abbas, and  Ziauddin}{Giant negative and positive Goos-H\"anchen shifts via 
Doppler broadening effect}{Las. Phys.}{29}{075201-6}{2019}

\bibitem{NP1}
\refa{R. Halir et al}{Ultra‐broadband nanophotonic beamsplitter using an anisotropic sub‐wavelength metamaterial}{Las. Phot. Rev.}{10}{1039-1046}{2016}

\bibitem{NP2}
\refa{L. Lambrechts, V. Ginis, J. Danckaert and P. Tassin} {Transformation optics for surface phenomena: Engineering the Goos-H\"anchen effect} {Phys. Rev. B}{95}{035427-5}{2017}

\bibitem{NP3}
\refa{M. Zoghi}{Goos–H\"anchen and Imbert–Fedorov shifts in a two-dimensional array of gold nanoparticles}
{J. Nanophot.}{12}{016021-9}{2018}

\bibitem{PC1}
\refa{Y.\,Dadoenkova et al}{Goos-H\"anchen effect in light transmission through biperiodic photonic-magnonic crystals}{Phys. Rev. A}{96}{043804-6}{2017}

\bibitem{PC2}
\refa{X. Wu}{Goos–H\"anchen shifts in tilted uniaxial crystals}{Opt. Comm.}{416}{181-184}{2018}

\bibitem{PC3}
\refa{I.\,V. Soboleva, V.\,V.  Moskalenko, and A.\,A. Fedyanin}{Giant Goos-H\"anchen Effect and Fano 
Resonance at Photonic Crystal Surfaces}{Phys. Rev. Lett.}{108}{123901}{2012}


\bibitem{DK18}
S. De Leo and R. Kraus, 
{\em  Incidence angles maximizing the Goos-Haenchen shift in seismic data analysis},
Pure App. Geoph. \textbf{175}, 2023-2044 (2018).



\bibitem{Exp1}
O. J. S. Santana, S. A. Carvalho, S. De Leo, and L. E. E. de Araujo,
{\em Weak measurement of the composite Goos-H\"anchen shift in the critical region},
 Opt. Lett. {\textbf{41}}, 3884-3887 (2016).

\bibitem{Exp2}
S. A. Carvalho, S. De Leo, J. O. A. Huguenin, M. Martino, L. Silva,
{\em Experimental evidence of laser power oscillations induced by the relative Fresnel 
(Goos-H\"anchen) phase},
Las. Phys. Lett. \textbf{16}, 065001-5 (2019).


\bibitem{Exp3}
O. J. S. Santana and L. E. E. de Araujo, 
{\em Direct measurement of the composite Goos-H\"anchen shift of an optical beam},
Opt. Lett. \textbf{43}, 4037-4040 (2018).







\bibitem{A13}
K.\,Y. Bliokh and A. Aiello,
{\em Goos-H\"anchen and Imbert-Fedorov beam shifts: an overview},
J. Opt. {\textbf{15}}, 014001-16 (2013).


\bibitem{M19}
S. De Leo and G.\,G. Maia,
{\em Lateral shifts and angular deviations
of Gaussian optical beams
reflected by and transmitted through
dielectric blocks: A tutorial review},
J. Mod. Opt. \textbf{66}, 2142-2194 (2019).







\bibitem{ADL}
\refb{S. Adler}{Quaternionic Quantum Mechanics and Quantum Fields}
{Oxford University Press}{1995}

\bibitem{del0}
\refa{S. De Leo and G. Ducati}{Quaternionic differential operators}{J. Math. Phys.}{42}{2236-2265}{2001}

\bibitem{del1}
\refa{S. De Leo, G. Ducati and C.\,C. Nishi}{Quaternionic potentials in non-relativistic quantum mechanics}{J. Phys. A}{35}{5411–5426}{2002}

\bibitem{del2}
\refa{S. De Leo and G. Ducati}{Quaternionic bound states}{J. Phys. A}{38}
{3443–3454}{2005}

\bibitem{del3}
\refa{S. De Leo, G. Ducati and T. Madureira}{Analytic plane wave solutions for the quaternionic potential step}{J. Math. Phys.}{47}{082106-15}{2006}

\bibitem{del4}
\refa{S. De Leo and G. Ducati}{Quaternionic diffusion by a potential step}{J. Math. Phys.}{47}{102104-9}{2006}

\bibitem{del5}
\refa{S. De Leo and G. Ducati}{Quaternionic wave packets}{J. Math. Phys.}{48}{052111-10}{2007}



\bibitem{a08}
\refa{T. Jiang and L. Chen}{An algebraic method for Schr\"odinger equations in quaternionic quantum mechanics}{Comp. Phys. Comm.}{178}{795-799}{2008}


\bibitem{a09}
\refa{F. Masillo, G. Scolarici, and  S. Sozzo}{Proper versus improper mixtures: Toward a quaternionic quantum mechanics}{Theor. Math. Phys.}{160}{1006-1013}{2009}



\bibitem{del6}
\refa{S. De Leo and G. Ducati}{Delay time in quaternionic quantum mechanics}{J. Math. Phys.}{53}{022102-8}{2012}

\bibitem{del7}
\refa{S. De Leo and S. Giardino}{Dirac solutions for quaternionic potentials}{J. Math. Phys.}{55}{022301-10}{2014}

\bibitem{a14}
\refa{T. Jiang, Z.Jianga, and  S. Ling}{An algebraic method for quaternion and complex Least Squares coneigen-problem in quantum mechanics}{App. Math. Comp.}{249}{222-228}{2014}

\bibitem{del8}
\refa{S. De Leo, G. Ducati and S. Giradino}{Quaternionic Dirac scattering}{J. Phys. Math.}{6}{1000130-6}{2015}




\bibitem{a16}
\refa{H. Sobhani and H. Hassanabadi}{Scattering in quantum mechanics under quaternionic Dirac delta potential}{Can. J. Phys.}{94}{262-266}{2016}


\bibitem{a17}
\refa{H. Hassanabadi, H. Sobhani, and   A. Banerjee}{Relativistic scattering of fermions in quaternionic quantum mechanics}{Eur. Phys. J. C}{77}{581-5}{2017}


\bibitem{a18}
\refa{J. Gantner}{On the equivalence of complex and quaternionic quantum mechanics}{Math. Found.}{5}{357-390}{2018}


\bibitem{del9}
\refa{S. De Leo, C.\,A.\,A. Almeida and  G. Ducati}{Quaternionic perturbation theory}{Eur. Phys. J. Plus}{134}{113-125}{2019}























\bibitem{SQP}
\refa{S. De Leo and G.C. Ducati}{The Snell law for quaternionic potentials}{J. Math. Phys}{54}{122109}{2013}








\bibitem{cohen}
\refb{C. C. Tannoudji, B. Diu and F. Lal\''oe}{Quantum Mechanics}{Wiley, Paris}{1977}


\bibitem{saleh}
B.\,E.\,A. Saleh and M.\,C. Teich,
{\em Fundamentals of Photonics}  (Wiley \& Sons,  2007).

\bibitem{born}
 M. Born and E. Wolf,
{\em Principles of optics} (Cambridge University Press, 1999).


\bibitem{spm}
\refa{S. Carvalho and S. De Leo}{The use of the stationary phase method as a mathematical tool to determine the path of optical beams}{Am. J. Phys.}{83}{249-255}{2015}


\bibitem{inf1}
\refa{M. Ara\'ujo, S. Carvalho, and S. De Leo}{The frequency crossover for the Goos–H\"anchen shift}{J.  Mod. Opt.}{60}{1772-1780}{2013}

\bibitem{inf2}
\refa{M. Ara\'ujo, S. De Leo, and G. Maia}{Closed form expression for the Goos-H\"anchen lateral displacement}
{Phys. Rev. A}{93}{023801-9}{2016}

\bibitem{inf3}
\refa{M. Ara\'ujo, S. De Leo, and G. Maia}{Oscillatory behavior of light in the composite Goos-H\"nchen shift}
{Phys. Rev. A}{95}{053836-9}{2017}


\bibitem{expq1}
\refa{A. Peres}{Proposed test for complex versus quaternion quantum theory}
{Phys. Rev. Lett.}{42}{683-686}{1979}

\bibitem{expq2}
\refa{H. Kaiser, E.  George,and S. Werner}{Neutron interferometric search for quaternions in quantum mechanics}
{Phys. Rev. A}{42}{2276-2279}{1984}

\bibitem{nature}
\refa{L.\,M. Procopio, et al.}{Single-photon test of hyper-complex quantum
theories using a metamaterial}
{Nat. Comm.}{8}{15044-9}{1984}


\bibitem{wm}
\refa{M. Araújo, S. De Leo, and G. Maia}{Optimizing weak measurements to detect angular deviations}
{Ann. der Physik}{529}{1600357-20}{2017}




\end{thebibliography}
\end{document}